# Soundtracks of Our Lives: How Age Influences Musical Preferences

Arsen Matej Golubovikj
matej.golubovikj@famnit.upr.si
University of Primorska
Koper, Slovenia

Bruce Ferwerda
bruce.ferwerda@ju.se
Jönköping University
Jönköping, Sweden

Alan Said
alansaid@acm.org
University of Gothenburg
Gothenburg, Sweden

Marko Tkalčič
marko.tkalcic@famnit.upr.si
University of Primorska
Koper, Slovenia

## ABSTRACT

The majority of research in recommender systems, be it algorithmic improvements, context-awareness, explainability, or other areas, evaluates these systems on datasets that capture user interaction over a relatively limited time span. However, recommender systems can very well be used continuously for extended time. Similarly so, user behavior may evolve over that extended time. Although media studies and psychology offer a wealth of research on the evolution of user preferences and behavior as individuals age, there has been scant research in this regard within the realm of user modeling and recommender systems. In this study, we investigate the evolution of user preferences and behavior using the LFM-2b dataset, which, to our knowledge, is the only dataset that encompasses a sufficiently extensive time frame to permit real longitudinal studies and includes age information about its users. We identify specific usage and taste preferences directly related to the age of the user, i.e., while younger users tend to listen broadly to contemporary popular music, older users have more elaborate and personalized listening habits. The findings yield important insights that open new directions for research in recommender systems, providing guidance for future efforts.

**ACM Reference Format:**
Arsen Matej Golubovikj, Bruce Ferwerda, Alan Said, and Marko Tkalčič. 2025. Soundtracks of Our Lives: How Age Influences Musical Preferences. In *Proceedings of Adjunct Proceedings of the 33rd ACM Conference on User Modeling, Adaptation and Personalization (UMAP Adjunct '25)*. ACM, New York, NY, USA, 5 pages. https://doi.org/10.1145/3708319.3733673

## 1 INTRODUCTION

Modeling user preferences and behaviors in music recommender systems is a complex domain explored in the literature [12, 16, 17]. While traditional studies often assume static user preferences over time, recent trends emphasize modeling transitional phases in user interactions and addressing challenges like the new user problem [19]. A notable perspective on this topic is the context modeling framework, exemplified by Koenigstein et al. [9], who analyzed users' weekly preference fluctuations alongside the declining popularity of songs over specific intervals. However, these approaches often focused on short time frames, typically weeks to months, rather than years or decades. Meanwhile, research in music psychology highlights that music preferences naturally evolve across different stages of age and life [2–4, 6, 7].

To date, no research has explored long-term changes in user preferences to inform recommender algorithms. For instance, while Ferwerda et al. [5] examined music behaviors across age groups, they did not track individual behavior over time and focused only on genre-based preferences.

This work bridges the gap by analyzing longitudinal shifts in preferences and behaviors of Last.fm[1] users by age. Using the 15-year LFM-2b dataset [14] (2005–2020), we address the following research questions:

**RQ1** How do user preferences and behavior on a music recommendation platform evolve as users age?
**RQ2** How can we model user preference changes based on a longitudinal analysis of behavior change?

## 2 RELATED WORK

This study bridges the gap between music psychology, which shows that preferences evolve over decades, and the limited research in user modeling and recommender systems that explores age-related changes in music preferences.

Holbrook and Schindler [8] examined how consumers' affinity for popular music changes with age, finding an inverted U-shaped curve with preferences peaking at a median age of 23.5 years and declining thereafter. Subsequent studies, such as Hemming [7] and Kopiez et al. [10], have confirmed similar patterns. Harrison and Ryan [6] focused on shifts in preferences among older adults compared to other age groups, noting that the range of favored genres expands in early adulthood, stabilizes around age 55, and contracts afterward. These findings were further supported by Ma [11] in an American sociocultural context. Bonneville-Roussy et al. [3] introduced the Music Preferences in Adulthood Model (MPAM), highlighting psychological factors like social influences and individual differences that shape preferences as listeners age. They observed that music engagement decreases with age and preferences evolve over time.

---
[1]https://www.last.fm/

Conversely, research in user modeling and recommender systems primarily emphasizes early development. Schedl and Bauer [13] examined the musical preferences of children and adolescents (ages 0–18), highlighting differences between younger and older individuals and suggesting that treating children as a distinct group could improve recommendations. Spear et al. [18] studied children aged 6–17, categorizing them into grade schoolers (ages 6–11), middle schoolers (ages 12–14), and high schoolers (ages 15–17). They found that children's tastes are diverse and increasingly divergent as they age.

Revisiting our research questions, **RQ1** builds on works like Holbrook and Schindler [8] and Kopiez et al. [10], which examine the evolution of music preferences to study psychological aspects. Our study, however, focuses on preference evolution from a recommender systems perspective, defining the scope of our analysis and contribution.

For **RQ2**, we draw inspiration from these works, particularly Kopiez et al. [10], but integrate a recommendation perspective, linking music recommendation [16] and user modeling [1] to a longitudinal analysis of music preference evolution.

## 3 ANALYZING LONGITUDINAL MUSIC PREFERENCES

We conducted a longitudinal study on the evolution of musical preferences and listening behavior using nearly 15 years of LFM-2b dataset listening events. Proxy measures include playcounts, unique tracks, unique artists, and diversity. Details on the data and measures are provided below.

### 3.1 Data

The LFM-2b dataset contains two billion listening events (LEs) from 120,322 users on Last.fm (February 2005–March 2020), recording timestamps, user IDs, track names, and artist names. Age data is available for a subset of users. Our analysis focuses on 42,883 users with plausible age data and consumption patterns meeting post-cleaning criteria (detailed below).

For this study, we refined the users' age parameter. The LFM-2b dataset originally includes a static age, reflecting users' age at data retrieval (between January 2013 and August 2014). To track age across the dataset's 15-year span (2005–2020), we adjusted each user's age based on the timing of their listening events. Using October 31, 2013, as the reference point $r_p$, we calculated the adjusted age $a_e$ for each event as $a_e = a + \Delta_y(t_e, r_p)$, where $a$ is the static age, and $\Delta_y(t_e, r_p)$ is the year difference between the event ($t_e$) and the reference point ($r_p$). This method minimizes age discrepancies to about ±1 year.

We restricted the age range to 10–64 years, categorizing users into fixed 5-year intervals: 10–14, 15–19, ..., 60–64. This interval width balances granularity and group size while maintaining consistent age diversity within groups. Inactive users and outliers were removed based on LEs for each group. Specifically, inactive users were excluded if their LEs fell below the 20th percentile. Whereas outliers were identified using the interquartile range (IQR): users with LEs below $Q1 - 1.5 \cdot IQR$ or above $Q3 + 1.5 \cdot IQR$ were omitted. Table 1 shows the number of users in each age group post-cleaning, noting that age adjustments allow users to appear in multiple groups. The cleaned dataset includes 42,883 users, 542,027,152 LEs, and 1,033,284 tracks with release years obtained via the Spotify API[2]. A comparison of the full and cleaned datasets is provided in Table 1.

|  | Listening events | Users | Tracks |
|---|---|---|---|
| **LFM 2b** | 2,014,164,872 | 120,322 | 50,813,373 |
| **Our subset** | 542,027,152 | 42,883 | 1,033,284 |

Table 1: Number of Users, LEs, and tracks in the original LFM-2b dataset and in the filtered subsets we used.

### 3.2 Proxy Measures

To track changes in music preferences and behaviors, we defined proxy measures grouped into *yearly consumption statistics* and *diversity measures*. Yearly consumption statistics include total playcounts ($P$), the number of unique tracks ($T$), and the number of unique artists ($A$) a user engaged with annually. Using $P$, $T$, and $A$, we assessed diversity based on the definition by Schedl and Hauger [15], calculated as $D_{\tau,u} = \frac{|T_u|}{P_u}$, the ratio of unique tracks ($|T_u|$) to playcounts ($P_u$). Following Spear et al. [18], we inverted this measure, placing diversity on a [0,1] scale, where 1 represents maximum diversity.

We expanded the concept of diversity with *artist-based diversity*, defined as $D_{\alpha,u} = \frac{|A_u|}{|T_u|}$, where $|A_u|$ and $|T_u|$ are the number of unique artists and tracks a user $u$ listens to, respectively. This metric reflects the range of a user's preferences across artists and tracks: concentrated playcounts on a few artists indicate low diversity, while distributed playcounts across many artists suggest high diversity.

### 3.3 Insights

We report changes in music preferences and streaming behavior with age using the proxy measures from Section 3.2. Findings are based on median values, chosen over the mean used in prior work [13, 18] due to distribution skewness.

Fig. 1 shows the median annual playcounts ($P$), unique tracks ($T$), and unique artists ($A$) by user age. Playcounts and unique tracks rise from age 10 ($P = 826$, $T = 281$), peaking at age 19 ($P = 2318$) and age 22 ($T = 805$), respectively, before declining to age 58 ($P = 560$) and age 54 ($T = 410$), respectively. Afterward, playcounts increase slightly, while unique tracks plateau. Unique artists ($A$) increase sharply from age 11 ($A = 65$) to 24 ($A = 178$), then rise gradually until age 38 ($A = 211$), followed by a decline to age 55 ($A = 145$) before stabilizing, with slight deviations at ages 60 ($A = 187$) and 64 ($A = 210$).

---

[2] https://developer.spotify.com/documentation/web-api

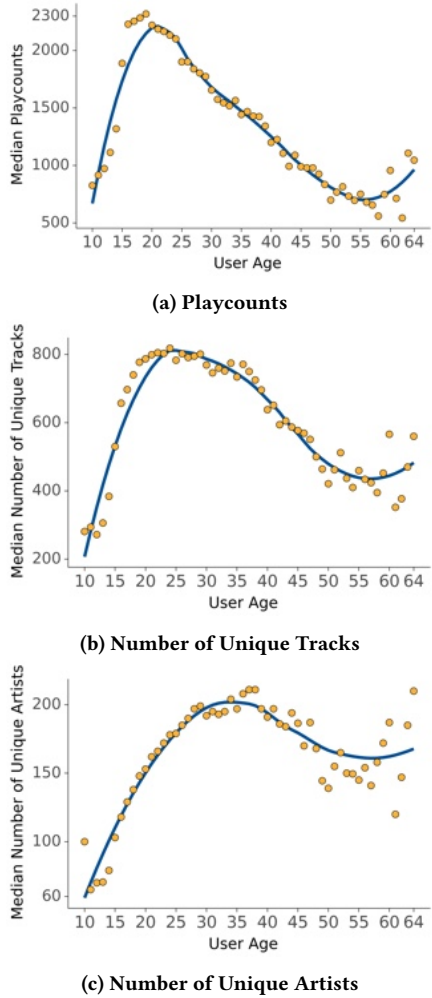

(a) Playcounts

(b) Number of Unique Tracks

(c) Number of Unique Artists

Figure 1: Median (a) playcounts, (b) number of unique artists, and (c) number of unique tracks users played per age. The exact median values for each age are depicted as dots, and a regression line is fitted, depicted in black, to show the overall trend of the values.

Fig. 2 shows median track-based diversity ($D_\tau$) and artist-based diversity ($D_\alpha$) by age. Both metrics start low at younger ages and increase as users mature, with $D_\tau$ rising faster than $D_\alpha$. Diversity grows significantly between ages 15 ($D_\tau$ = 0.316, $D_\alpha$ = 0.189) and 40 ($D_\tau$ = 0.599, $D_\alpha$ = 0.328), reflecting broader listening preferences with age. While both metrics show similar trends, differences are more pronounced in $D_\tau$, which we prioritize as the main diversity metric for further analysis. These findings address **RQ1**: How do user preferences and behavior on a music recommendation platform evolve with age?

We analyzed diversity variation within age groups using the Gini index, as shown in Fig. 3. The Gini index ([0, 1]) measures distribution spread (0 = equal values, 1 = highly spread). While diversity increases with age, variation is highest at ages 10–14 (*Gini* = 0.336 at age 10 and 0.325 at age 14) and declines sharply between ages 15 (*Gini* = 0.308) and 40 (*Gini* = 0.169), stabilizing after 40 with minor fluctuations beyond 60. This suggests younger users exhibit greater diversity in consumption habits, converging as they age.

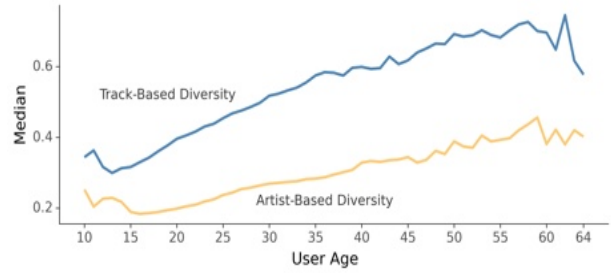

Figure 2: Median track-based diversity $D_\tau$ and artist-based diversity $D_\alpha$ per user age. Where $D_\tau$ and $D_\alpha$ are our diversity metrics in the range of [0,1] (defined in Section 3.2), with higher values indicating more diversity.

We also examined diversity in the context of track release years. Fig. 4 shows median $D_\tau$ by release year across age groups. All groups exhibit peak diversity ($D_\tau$ = 1) for tracks released before 1960, reflecting broad tastes for older tracks. Ages 10–14 show high diversity for pre-70s tracks ($D_\tau \approx 1$), moderate diversity for tracks from the 70s–90s ($D_\tau \approx 0.61$), and a decline for newer releases ($D_\tau$ = 0.47 to 0.23). Ages 15–44 follow a similar pattern, with diversity increasing slightly and becoming less distinct. Ages 45–64 show more uniform diversity across release years ($D_\tau \approx 0.8$). This aligns with Gini index findings, indicating greater differentiation in youth and a shift toward uniformity with age.

We further explored the temporal relationship between release year, age, and preference using the song-specific age (SSA) metric, introduced by Holbrook and Schindler [8]. SSA represents the user's age at a song's release, calculated as *user_age* − *song_age*. Median log-normalized playcounts (base 10, scaled between 0 and 3.903) were computed for each SSA and grouped by age categories (Fig. 5). Log normalization addressed two issues: (i) differing playcount magnitudes across ages (Fig. 1) made visual comparisons challenging, and normalization aligned values for comparability; (ii) skewed distributions were adjusted, bringing tail values closer

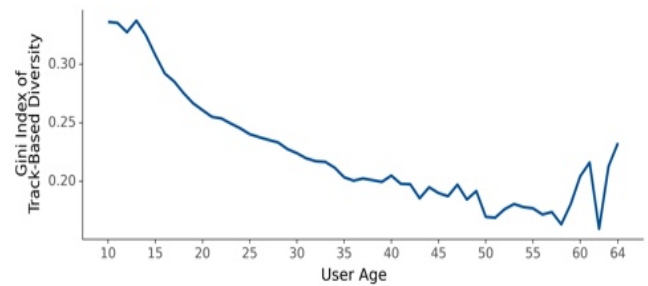

Figure 3: The Gini index of track-based diversity, based on the user's age, i.e., the diversity of diversity. The lower values mean less dispersion (values are more similar, less diversity)

to the head for clearer visibility. Base 10 was chosen for optimal handling of high playcounts.

SSA represents the user's age when a track debuted, with negative values indicating the track was released before their birth. Results show a strong preference for contemporary music, particularly among younger users. Playcounts peak when SSA matches the user's age, with higher peaks for younger groups ($P = 110$ at ages 10–14, $P = 186$ at ages 15–19) and diminishing peaks as users age ($P = 155$ at 20–24, $P = 89$ at 30–34, $P = 52$ at 40–44, $P = 23$ at 55–59). Tail playcounts remain stable ($P \approx 10$). After age 45, a bimodal distribution emerges, with a peak for current releases ($P \approx 33$) and a secondary peak for tracks from users' adolescence (SSA of 15–20 years, $P \approx 13$).

The decline in contemporary music playcounts with age explains the rise in diversity (Fig. 4) and reduced Gini index (Fig. 3). The bimodal pattern in older age groups indicates a shift toward singular listening preferences. These findings address **RQ2**: How can we model user preference changes based on longitudinal behavior analysis?

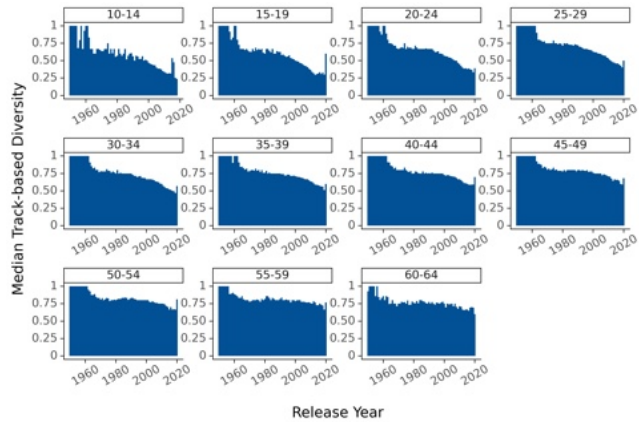

**Figure 4: The median track-based diversity $D_{\tau,u}$ in each age group per release year. Higher values indicate more diversity.**

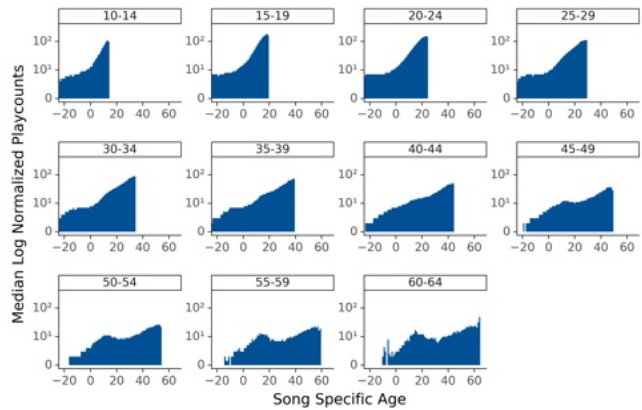

**Figure 5: Median Log normalized playcounts in each age group per song specific age.**

## 4 DISCUSSION AND IMPLICATIONS FOR RECOMMENDER SYSTEMS

Our analysis reveals that as users age, intra-user diversity increases (Fig. 2), while inter-user diversity decreases (Fig. 3). Younger users exhibit lower individual diversity but greater variation as a group. Recommender systems should adjust diversity and personalization based on user age: younger users benefit from low personalization and high diversity, while older users prefer high personalization and low diversity. This aligns with studies showing teenagers and young adults are more engaged with popular culture than older adults [2–4, 6, 7] and builds on findings about genre-based diversity by age [5].

SSA analysis provides further insights. Younger listeners prefer songs with SSA matching their age, favoring new releases while exploring older music. This pattern persists until around age 40, after which preferences shift to music from their youth (SSA 10–20 years). Beyond 40, current music consumption declines, and nostalgia-driven listening dominates. Recommender systems should: (1) tailor recommendations for younger users to highlight contemporary music while encouraging exploration of older tracks, (2) provide middle-aged listeners a balance between contemporary and nostalgic preferences, and (3) focus recommendations for older users on music aligned with their refined tastes and past preferences.

Finally, algorithmic confounding, the influence of recommendation algorithms on listening behaviors over time, can unintentionally shape diversity across age groups. Younger users may receive varied recommendations due to broader initial interests, fostering diversity, while older users might encounter less variety, reinforcing existing preferences. This feedback loop significantly impacts the evolution of music preferences. Recommender systems should adopt adaptive strategies to align with the changing diversity preferences of different age groups.

### 4.1 Limitations

This research has several limitations. First, unreliable data, such as improbable user ages, raises concerns about the accuracy of the age variable, potentially biasing conclusions. Second, the dataset is limited to users active in 2013–2014, excluding new accounts post-2014 and favoring long-term users, which skews the analysis and limits applicability to modern contexts. Additionally, tracks are unevenly distributed by release year, introducing biases that may misrepresent the popularity or significance of certain periods. Finally, the study employs a basic diversity metric that may oversimplify musical diversity, potentially missing nuances. A more sophisticated metric could yield more accurate and insightful results.

## 5 CONCLUSION

This work examined how music preferences and behaviors evolve with age using a 15-year longitudinal analysis of the LFM-2b dataset. Findings indicate that as users age, preferences shift from highly diverse and non-personalized content to lower diversity but more personalized content. Younger listeners primarily favor recent releases, but this preference diminishes with age, transitioning to a mix of recent and nostalgic tracks. By age 40, users predominantly engage in reminiscence, listening to music from their adolescent years.

These insights highlight how recommender systems can better serve users of different ages. Younger listeners benefit from recommendations balancing contemporary and older tracks, while older users prefer personalized suggestions emphasizing nostalgic content. Algorithms that adjust diversity and personalization based on age can better meet the varied preferences across age groups.

## ACKNOWLEDGMENTS

The work presented here was supported by the University of Primorska grant CogniCom and the Slovenian Research and Innovation Agency through the grant N2-0354. The authors would also like to acknowledge the support of Markus Schedl, Alessandro Melchiorre, and Stefan Brandl of the Johannes Kepler University Linz (Austria) in dealing with the LFM-2b dataset.